\documentclass[aps,pre,reprint,twocolumn,superscriptaddress,longbibliography,floatfix]{revtex4-2}

\usepackage{graphicx}
\usepackage{amssymb}
\usepackage{xcolor}
\usepackage{mathtools}
\usepackage{array}
\usepackage{bm}
\usepackage[colorlinks=true, citecolor=blue, linkcolor=blue, urlcolor=blue]{hyperref}
\usepackage{booktabs}

\begin{document}

\title{Feedback-mediated circulation and persistence of stochastic fluctuations in gene regulatory circuits}

\author{Nashita Rahman}
\email{nashita.rahman@jcbose.ac.in}
\affiliation{Department of Chemical Sciences, Bose Institute, EN 80, Sector V, Bidhan Nagar, Kolkata 700091, India}

\author{Mintu Nandi}
\email{mintunandi@ubi.s.u-tokyo.ac.jp}
\thanks{Corresponding author}
\affiliation{Universal Biology Institute, The University of Tokyo, 7-3-1 Hongo, Bunkyo-ku, Tokyo 113-0033, Japan}

\author{Sudip Chattopadhyay}
\email{sudip@chem.iiests.ac.in}
\affiliation{Department of Chemistry, Indian Institute of Engineering Science and Technology, Shibpur, Howrah 711103, India}

\author{Suman K Banik}
\email{skbanik@jcbose.ac.in}
\thanks{Corresponding author}
\affiliation{Department of Chemical Sciences, Bose Institute, EN 80, Sector V, Bidhan Nagar, Kolkata 700091, India}

\date{\today}

\begin{abstract}
Feedback plays a significant role in biochemical networks that govern a multitude of cellular functions, including development, adaptation, and homeostasis. Yet, how feedback topology controls stochastic fluctuations remains incompletely understood. Here, we develop a theoretical framework for two-node feedback motifs composed of activating and repressive regulatory interactions between two transcription factors. Under the linear noise approximation, we identify a feedback-driven contribution to node-wise fluctuations, termed cyclic noise, that arises specifically from loop closure. Cyclic noise is the component of fluctuations that circulates through the regulatory circuit. Its sign and magnitude distinguish whether feedback amplifies or attenuates node-wise fluctuations. We further show that feedback-mediated noise circulation leaves a temporal signature in the decay of steady-state autocorrelation, revealing how loop closure modifies the persistence of fluctuations. We thus provide a minimal framework for understanding how feedback architecture regulates both the magnitude and the temporal persistence of noise in gene regulatory circuits.
\end{abstract}

\maketitle


\section{Introduction}

\begin{figure*}[!t]
    \includegraphics[width=1\linewidth]{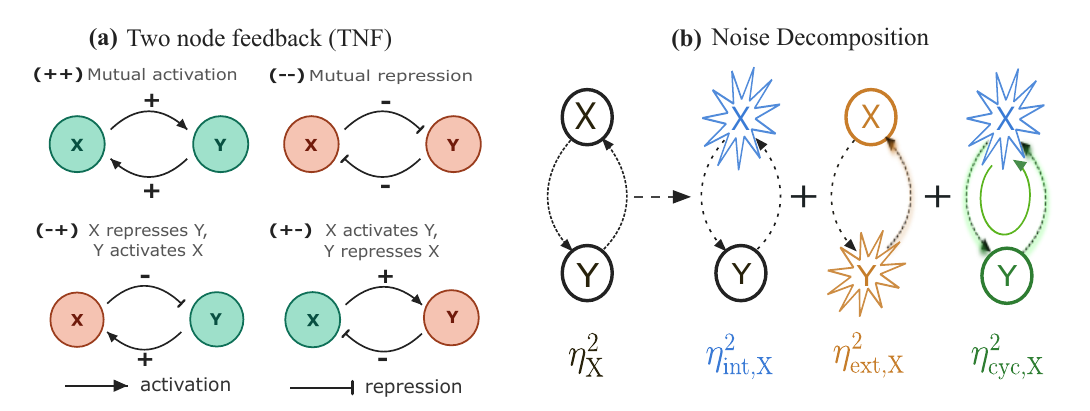}
    \caption{
    Schematic representation of two-node feedback motifs and feedback-dependent noise decomposition. 
    (a) The four possible TNF motifs formed by activating and repressive regulations between two TFs, $X$ and $Y$. 
    (b) Noise decomposition for TF $X$ in a TNF motif. In addition to the conventional intrinsic and extrinsic noise components, feedback loop closure generates a cyclic noise component that represents the returned contribution of fluctuations circulating through the regulatory loop. The corresponding decomposition for TF $Y$ follows analogously.
    }
    \label{fig1}
\end{figure*}

Noise is ubiquitous in biological systems as biochemical reactions are probabilistic and molecular species operate in fluctuating cellular environments \cite{Elowitz2002}. To maintain reliable function under such variability, cells use complex regulatory networks composed of many interacting biochemical components \cite{Alon2006}. The networks are often organized into recurring regulatory motifs, among which feedback motifs play an important role in decision-making, adaptation, and noise control in biochemical networks \cite{Becskei2000, Angeli2004}. Autoregulation represents the simplest feedback architecture, in which a transcription factor (TF) regulates its own expression. A two-node feedback (TNF) motif (see Fig.~\ref{fig1}a) provides the next level of feedback organization, where two TFs regulate each other through activating or repressive interactions.


Genetic feedback motifs can give rise to rich dynamical behaviors, including multistability, switching, oscillations, and stimulus retention, depending on the nonlinearity and cooperativity of the regulatory functions \cite{Angeli2004, Cherry2000}. Feedback loops have therefore been studied extensively from both theoretical and experimental perspectives \cite{Lestas2010, Rhee2014, Tkacik2009, Smolen2009, Chang2010, Hinczewski2016, Jia2017}. For example, Chang \textit{et al.} examined memory through two interacting positive feedback loops and showed that nonlinear coupling between the loops can produce an ultrasensitive bistable response \cite{Chang2010}. Jia \textit{et al.} showed that steady-state protein-number distributions can provide diagnostic signatures of distinct feedback mechanisms \cite{Jia2017}. Hinczewski \textit{et al.} quantified how negative feedback mitigates noise using the principle of optimal linear filters \cite{Hinczewski2016}. Smolen \textit{et al.} studied how coupled fast and slow feedback loops improve memory robustness in the presence of fluctuations \cite{Smolen2009}. These studies show that feedback significantly influences stochastic dynamics, yet a comprehensive analysis of feedback-mediated noise propagation in minimal TNF motifs remains underdeveloped.

In control theory, feedback loops are known to modify fluctuations by either amplifying or damping them \cite{Cosentino2011}. In biological systems, however, gene-expression noise can propagate across different levels of network organization and influence essential cellular processes \cite{Tsimring2014}. A classical method for understanding gene expression variability is noise decomposition, which separates total variability into intrinsic and extrinsic components \cite{Elowitz2002, Swain2002}. This conventional approach is effective for identifying local and external sources of fluctuations, but it is not complete for closed feedback motifs. In a feedback loop, fluctuations generated at one node can affect another node and return to the original node through the closed regulatory path. This leads to a central question: Can the contribution generated by the closure of the feedback loop be isolated as a distinct component of noise?
While feedback loop gains have been introduced to quantify feedback efficiency \cite{Kobayashi2016}, an explicit additive loop-closure contribution to node-wise noise and its temporal signature has not been characterized.

To address this question, we present a theoretical framework for TNF motifs to examine how the closure of feedback loops influences noise propagation. We consider four distinct TNF topologies based on the regulatory interactions between two nodes: mutual activation (represented as $++$), mutual repression ($--$), and two mixed-sign motifs ($-+$ and $+-$) (see Fig.~\ref{fig1}a). Our focus is on the monostable steady-state regime, where we introduce a feedback-specific contribution to node-wise noise. This contribution accounts for fluctuations generated by feedback loop closure and enables us to assess whether a particular feedback topology amplifies or attenuates them. Additionally, we investigate whether this feedback-mediated noise circulation leaves a temporal signature in the autocorrelation of steady-state fluctuations. Overall, this study explores how the architecture of feedback loops controls both the magnitude and temporal persistence of stochastic fluctuations in two-node gene regulatory circuits.


\section{The Model}

The TNF motif is modeled within the stochastic framework using the Langevin formalism. The dynamics of the TFs in this motif are described by the following stochastic differential equation \cite{Bintu2005, Ziv2007, Tkacik2008a, Tkacik2008b, deRonde2012, Walczak2012},
\begin{equation}
    \frac{d \bm{n}}{dt} = \bm{\mathcal{F}}(\bm{n}) - \mathbf{\mathcal{B}} \bm{n} + \sqrt{2} \mathbf{\Delta} \bm{\xi}(t),
    \label{eq1}
\end{equation}

\noindent where $\bm{n}=(n_X,n_Y)^\top \equiv \left( x, y \right)^\top$ denotes the vector of copy numbers (state space) of the two TFs $X$ and $Y$. These copy numbers are expressed in molecules/V, where V refers to the cellular volume. The regulatory structure of each TNF is encoded in the synthesis-rate vector $\bm{\mathcal{F}}(\bm{n})=\left( f_X(\bm{n}), f_Y(\bm{n}) \right)^\top$, where the functions $f_X$ and $f_Y$ are described using Hill-type forms for activation and repression (see Table~\ref{t1}). The degradation of each TF is assumed to follow first-order kinetics and, therefore, $\mathbf{\mathcal{B}}={\rm diag}\!\left( \beta_X, \beta_Y \right)$ represents the matrix of degradation rate constants. In the stochastic description, $\bm{\xi}(t)=\left( \xi_X, \xi_Y \right)^\top$ accounts for the Gaussian white noise vector satisfying $\langle \xi_N(t) \rangle=0$ and $\langle \xi_N(t) \xi_M(t^\prime) \rangle=\delta_{NM}\delta(t-t^\prime)$ with $N, M \in \{ X, Y \}$. Here, $\langle \cdots \rangle$ denotes ensemble averaging over many realizations. The statistical properties of noise imply that the noise processes associated with the two TFs are independent and temporally uncorrelated \cite{Kampen2007, Elf2003, Paulsson2004, Swain2004, Tanase2006, deRonde2010}. Note that the stochastic term in Eq.~(\ref{eq1}) contains a noise amplitude matrix $\mathbf{\Delta}={\rm diag}\!\left( \Delta_X, \Delta_Y \right)$. The corresponding diffusion matrix, analogous to that appearing in the Fokker-Planck description, is given by $\bm{\mathcal{D}}=\mathbf{\Delta}\mathbf{\Delta}^\top={\rm diag}\!\left( \mathcal{D}_X, \mathcal{D}_Y \right)$. Under the linear noise approximation (LNA), the diffusion coefficients at steady-state are given by $\mathcal{D}_N = \left[ f_N(\langle\bm{n}\rangle) + \beta_N \langle n_N\rangle \right]/2$ \cite{Swain2004}, where $f_N(\langle\bm{n}\rangle)$ refers to the production rate of TF $N$ evaluated at steady-state, with corresponding state space $\langle\bm{n}\rangle=(\langle n_X\rangle, \langle n_Y\rangle)^\top \equiv \left( \langle x\rangle, \langle y\rangle \right)^\top$ representing the vector of steady-state copy numbers.

Within the linear noise approximation (LNA), Eq.~(\ref{eq1}) can be solved analytically in terms of the steady-state covariance matrix by using the Lyapunov equation,
\begin{equation}
    \mathbf{J} \mathbf{\Sigma} + \mathbf{\Sigma} \mathbf{J}^\top + 2\bm{\mathcal{D}} = 0,
    \label{eq2}
\end{equation}

\noindent where, $\mathbf{\Sigma}$ is the covariance matrix of fluctuations in copy numbers and $\mathbf{J}$ is the Jacobian matrix of the deterministic drift vector $\bm{\mathcal{G}}(\bm{n})=\bm{\mathcal{F}}(\bm{n}) - \mathbf{\mathcal{B}} \bm{n}$ evaluated at steady-state. We note that the drift vector can also be represented as $\bm{\mathcal{G}}(\bm{n})=\left( \mathcal{G}_X(\bm{n}), \mathcal{G}_Y(\bm{n}) \right)^\top$. The elements of the steady-state Jacobian are thus defined as $J_{NM}=(\partial \mathcal{G}_N/\partial n_M)_{\bm{n}=\langle \bm{n}\rangle}$. Analytical expressions of the elements of the covariance matrix are derived from the Lyapunov equation in Appendix~\ref{a1}. From these expressions, we define noise associated with each TF by the squared coefficient of variation (CV). For instance, noise in TF $N$ is $\eta_N^2=\sigma_N^2/\langle n_N\rangle^2$, where $\sigma_N^2$ stands for the variance in the copy number of TF $N$. Proper analysis of the noise in $X$ ($\eta_X^2$) and in Y ($\eta_Y^2$), based on noise decomposition, can be used to infer the noise propagation within the motif.


\begin{table*}[!t]
  \centering
  \caption{Synthesis and degradation propensities for the four TNF variants. Synthesis is described by Hill-type regulatory functions with a Hill coefficient of 1, and degradation follows first-order kinetics. The corresponding TNF topologies are illustrated in Fig.~\ref{fig1}a. Here, $\alpha_N$ is the synthesis rate constant with units of (molecules/V)min$^{-1}$, $K_{NM}$ is the dissociation constant of a TF $N$ from the promoter of gene $M$ with units of molecules/V, and $\beta_N$ is the degradation rate constant with units of min$^{-1}$. Note that V denotes the cellular volume.
  }
  \label{t1}
  \begin{tabular*}{\textwidth}{@{\extracolsep{\fill}} l cc cc} 
  \toprule
  \textbf{Two-node feedback variants}
      & \multicolumn{2}{c}{\textbf{Synthesis [$f_N(\bm{n})$]}}
      & \multicolumn{2}{c}{\textbf{Degradation [$\beta_N n_N$]}} \\
    \cmidrule(lr){2-3} \cmidrule(lr){4-5}
                       & \textbf{$x \rightarrow x+1$} & \textbf{$y \rightarrow y+1$}
                       & \textbf{$x \rightarrow x-1$} & \textbf{$y \rightarrow y-1$} \\
    \midrule
    Mutual activation ($++$) &  $\alpha_X \frac{y}{y+K_{YX}}$ & $\alpha_Y \frac{x}{x+K_{XY}}$ & $	\beta_X x$ &  $\beta_Y y$ \\
    \\
     Mutual repression ($--$) &  $\alpha_X \frac{K_{YX}}{y+K_{YX}}$ & $\alpha_Y \frac{K_{XY}}{x+K_{XY}}$ & $\beta_X x$ &  $\beta_Y y$ \\
     \\
     X represses Y and Y activates X ($-+$) & $\alpha_X \frac{y}{y+K_{YX}}$ & $\alpha_Y \frac{K_{XY}}{x+K_{XY}}$ & $\beta_X x$ &  $\beta_Y y$ \\
     \\
     X activates Y and Y represses X ($+-$) & $\alpha_X \frac{K_{YX}}{y+K_{YX}}$ & $\alpha_Y \frac{x}{x+K_{XY}}$ & $\beta_X x$ &  $\beta_Y y$ \\
    \bottomrule
  \end{tabular*}
\end{table*}


\section{Results and discussion}

\subsection{Emergence of feedback-mediated cyclic noise}

Noise propagation in gene-regulatory motifs can be intuitively decomposed into intrinsic and extrinsic components \cite{Paulsson2004, Elowitz2002}. From the perspective of gene regulation, intrinsic noise refers to the fluctuations caused by the birth-death kinetics of the TF itself, which is typically Poisson in nature \cite{Nandi2024, Nandi2024_a}. In contrast, the extrinsic noise associated with the TF arises from external factors affecting gene regulation, resulting in total noise that is super-Poisson in nature \cite{Nandi2024, Nandi2024_a}. 

We apply this noise decomposition principle to analyze the total noise of each TF in the context of TNF, revealing an additional noise component that extends beyond just intrinsic and extrinsic factors. In the following, we present the explicit framework for noise decomposition as applied to the TNF motifs shown in Fig.~\ref{fig1}b.
For the TF $N$, the decomposition can be shown as,
\begin{equation}
    \eta_N^2 = \eta_{\text{int}, N}^2 + \eta_{\text{ext}, N}^2 + \eta_{\text{cyc}, N}^2,
    \label{eq3}
\end{equation}

\noindent where the three terms on the right-hand side represent the intrinsic, extrinsic, and additional noise contributions, respectively. These contributions are given by (see Appendix~\ref{a2}),
\begin{eqnarray}
    \eta_{\text{int}, N}^2 &=& \frac{1}{\langle n_N \rangle}
    \label{eq4} \\
    \eta_{\text{ext}, N}^2 &=& \left[
    \frac{(f^\prime_{NM})^2 \langle n_M \rangle}{\beta_N (\beta_N + \beta_M)\langle n_N \rangle} \eta_{\text{int}, N}^2 \right] \frac{1}{1-\mathcal{H}}
    \label{eq5} 
    \\
    \eta_{\text{cyc}, N}^2 &=&
    \mathcal{T}_{MN} \frac{\mathcal{H}}{(1-\mathcal{H})} \eta_{int,N}^2,
    \label{eq6}
\end{eqnarray}

\noindent where $N \neq M$ holds. Here, $f^\prime_{NM}=[\partial f_N/\partial n_M]_{\bm{n}=\langle \bm{n}\rangle}$ describes the steady-state regulatory sensitivity for the regulation $M \dashrightarrow N$. We note that $\dashrightarrow$ represents a general notation for both activation ($\rightarrow$) and repression ($\dashv$). This sensitivity measures how strongly a small change in $M$ affects the steady-state production rate of $N$. In the following, we define and interpret the remaining quantities appearing in Eqs.~(\ref{eq4}-\ref{eq6}) sequentially.

Note that $\mathcal{H}$ in Eqs.~(\ref{eq5})~and~(\ref{eq6}) is referred to as ``feedback gain", and is defined as,
\begin{equation}
    \mathcal{H}= \frac{f^\prime_{XY} f^\prime_{YX}}{\beta_X \beta_Y}.
    \label{eq7}
\end{equation}

This dimensionless quantity characterizes the strength of feedback coupling within the TNF motif. Since \(\mathcal{H}\) is expressed as the product of the two regulatory sensitivities associated with the reciprocal regulations in the TNF motif, it serves as a compact measure of how strongly the two regulatory edges are coupled. Therefore, \(\mathcal{H} = 0\) indicates that closed feedback interaction is absent, while \(\mathcal{H} \neq 0\) signifies the presence of feedback and specifies its nature.

When the two regulatory edges have the same sign, as in the \((++)\) and \((--)\) TNF motifs (see Fig.~\ref{fig1}a), we find that \(\mathcal{H} > 0\). This implies that the feedback is reinforcing. In the finite-noise regime, \(\mathcal{H} < 1\), as can be seen from Eqs.~(\ref{eq5}) and~(\ref{eq6}). Furthermore, it is clear that at \(\mathcal{H} = 1\), the determinant of the Jacobian, \(\det(\mathbf{J})\), becomes zero, leading to a singular Jacobian and causing the variances to diverge to infinity. Thus, for reinforcing feedback motifs, values of \(\mathcal{H}\) that are closer to unity indicate stronger feedback coupling within the finite-noise regime.

In contrast, when the two regulatory edges have opposite signs, as in the \((+-)\) and \((-+)\) TNF motifs (see Fig.~\ref{fig1}a), we find that \(\mathcal{H} < 0\). This corresponds to opposing feedback, where larger negative values of \(\mathcal{H}\) indicate stronger opposing feedback coupling. It is worth noting that for the \((+-)\) and \((-+)\) TNF motifs, \(\mathcal{H}\) can take on any negative value, including \(\mathcal{H} = -1\), depending on the model parameters. In this scenario, the Jacobian and Eqs.~(\ref{eq5}) and~(\ref{eq6}) do not impose a corresponding finite-noise boundary at \(\mathcal{H} = -1\).

The limiting behaviors of \(\mathcal{H}\) for different TNF motifs are examined quantitatively by tuning system parameters in Appendix~Fig.~\ref{fig_a1}. Therefore, the feedback gain \(\mathcal{H}\) helps identify both the strength and the sign of feedback coupling, proving to be a useful quantity for interpreting node-wise noise propagation within TNF motifs. Additionally, the quantity $\mathcal{T}_{MN} = \beta_M/( \beta_N + \beta_M)$, appearing in Eq.~(\ref{eq6}), defines the dimensionless time-averaging factor.

With these definitions, the three terms in Eq.~(\ref{eq3}) can be interpreted separately. The first term, $\eta_{\text{int}, N}^2$, represents the intrinsic noise of TF $N$. It has the Poisson form given in Eq.~(\ref{eq4}) and therefore provides the noise baseline set by the birth-death kinetics of $N$. The second term, $\eta_{\text{ext}, N}^2$, represents the extrinsic noise of $N$ arising from its regulatory dependence on its cognate TF $M$, where $M \neq N$. To be precise, the origin of this extrinsic noise is in the intrinsic noise of $M$, which causes $N$ to fluctuate beyond its Poisson limit (see Fig.~\ref{fig1}b). The feedback dependence of the extrinsic term can be read from Eq.~(\ref{eq5}) as,
\begin{eqnarray*}
    \eta_{\text{ext}, N}^2 &=& \underbrace{\left[
    \frac{(f^\prime_{NM})^2 \langle n_M \rangle}{\beta_N (\beta_N + \beta_M)\langle n_N \rangle} \eta_{\text{int}, N}^2 \right]}_{\substack{\text{Extrinsic contribution due to } \\ \text{$M \dashrightarrow N$ regulation}}}
    \underbrace{\frac{1}{1-\mathcal{H}}}_{\substack{\text{Feedback-dependent} \\ \text{modification}}}.
\end{eqnarray*}

\noindent Thus, the first factor represents the contribution associated with the direct regulation $M \dashrightarrow N$, whereas the second factor accounts for the modification introduced by feedback closure.

The third term in Eq.~(\ref{eq3}), $\eta_{\text{cyc}, N}^2$, has a distinct origin and can be understood directly from its explicit form in Eq.~(\ref{eq6}). In essence, this additional noise can vanish in the absence of feedback, i.e., 
\begin{equation*}
    \lim_{\mathcal{H} \to 0} \eta_{\text{cyc}, N}^2 = 0.
\end{equation*}

\noindent However, for $(++)$ and $(--)$ TNF motifs, as $0<\mathcal{H}<1$, this noise contribution is positive and increases as $\mathcal{H}$ approaches unity, i.e.,
\begin{equation*}
    \lim_{\mathcal{H} \to 1^-} \eta_{\text{cyc}, N}^2 = +\infty.
\end{equation*}

\noindent In contrast, for $(+-)$ and $(-+)$ TNF motifs, as $\mathcal{H}<0$, this contribution becomes negative. For example,
\begin{equation*}
    \lim_{\mathcal{H} \to -1} \eta_{\text{cyc}, N}^2 = -\frac{1}{2} \mathcal{T}_{MN} \, \eta_{\text{int}, N}^2.
\end{equation*}

\noindent These limiting behaviors show that the third contribution is positive for reinforcing feedback [$(++)$ and $(--)$ TNF motifs] and negative for opposing feedback [$(+-)$ and $(-+)$ TNF motifs], while disappearing entirely in the absence of feedback.

These limits also clarify why this additional term is not simply another intrinsic or extrinsic contribution. Its proportionality to $\eta_{\text{int}, N}^2$ indicates that it is tied to fluctuations generated at the same node whose noise is being measured. However, its dependence on $\mathcal{H}$ shows that this returned contribution arises only when the feedback path is active. In addition, the factor $\mathcal{T}_{MN}$ accounts for time-averaging in the two-node system. Since the returned contribution passes through the other TF $M$, its contribution to the noise of $N$ is weighted by the time-averaging factor. In essence, Eq.~(\ref{eq6}) can be read as,
\begin{eqnarray*}
    \eta_{\text{cyc}, N}^2 &=&
    \underbrace{\mathcal{T}_{MN}}_{\substack{\text{Time-averaging} \\ \text{factor}}} 
    \underbrace{\frac{\mathcal{H}}{(1-\mathcal{H})}}_{\substack{\text{Feedback-gain} \\ \text{factor}}} 
    \underbrace{\eta_{\text{int},N}^2}_{\substack{\text{Intrinsic} \\ \text{noise}}}.
\end{eqnarray*}

\noindent This structure demonstrates that a fluctuation in TF \( N \) can influence its corresponding TF \( M \). This influence can then feed back to affect \( N \) again, creating a closed loop in noise propagation represented as \( N \dashrightarrow M \dashrightarrow N \). We refer to this phenomenon as cyclic noise. It is specifically the feedback-dependent component resulting from the loop closure and has no equivalent in an open regulatory cascade.


\begin{figure*}
    \centering
    \includegraphics[width=1.0\linewidth]{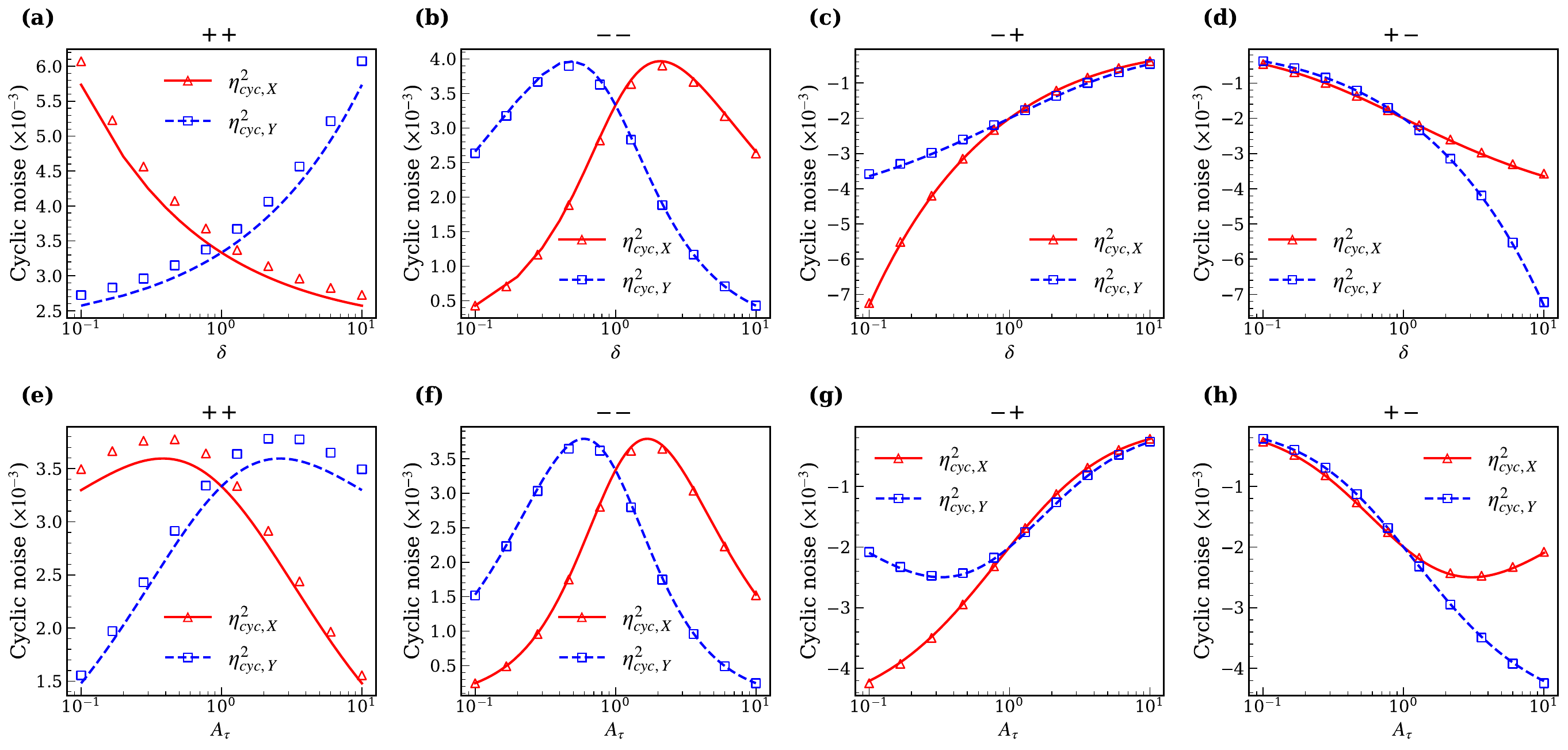}
    \caption{Dependence of cyclic noise on feedback-strength $\delta$ and timescale asymmetry $A_\tau$ in TNF motifs. 
    (a--d) Variation of cyclic noise with $\delta$, at fixed $A_\tau=1$. 
    (e--h) Variation of cyclic noise with $A_\tau$, at fixed $\delta=1$. 
    Panels (a,e), (b,f), (c,g), and (d,h) correspond to $(++)$, $(--)$, $(-+)$, and $(+-)$ TNF motifs, respectively.
    Solid and dashed lines represent theoretical predictions, whereas symbols denote stochastic simulation results averaged over $10^5$ trajectories \cite{Gillespie1976, Gillespie1977}. 
    The parameters are $\alpha_X=10$ (molecules/V)min$^{-1}$, $\alpha_Y=10$ (molecules/V)min$^{-1}$, $\beta_X=0.1\sqrt{A_\tau}$ min$^{-1}$, $\beta_Y=0.1/\sqrt{A_\tau}$ min$^{-1}$, $K_{XY}=50\sqrt{\delta}$ molecules/V, and $K_{YX}=50/\sqrt{\delta}$ molecules/V.
    }
    \label{fig2}
\end{figure*}

\subsection{Motif-specific nature of cyclic noise}

Given that \(\eta_{\text{cyc},N}^2 > 0\) indicates reinforcing feedback and \(\eta_{\text{cyc},N}^2 < 0\) signifies opposing feedback motifs, cyclic noise carries a significant physical meaning. Positive cyclic noise suggests that loop closure reinforces fluctuations at the node level, thereby increasing the overall noise of TF \(N\). In contrast, negative cyclic noise does not indicate negative physical noise; instead, it reflects a suppressive effect caused by opposing feedback, where the cyclic fluctuations counteract the noise accumulated at TF \(N\). This distinction provides a direct method for characterizing how the topology of feedback influences the propagation of stochastic noise in gene-regulatory motifs.

To quantify the nature of cyclic noise in a motif-specific manner, we introduce two parameters: feedback strength (\(\delta\)) and time-scale asymmetry (\(A_{\tau}\)). The feedback strength is defined as the ratio of the binding constants for the regulations \(X \dashrightarrow Y\) and \(Y \dashrightarrow X\), specifically \(\delta = K_{XY}/K_{YX}\). When \(\delta = 1\), both regulatory arms have identical binding thresholds. However, when \(\delta \neq 1\), an asymmetry is introduced between the two regulatory directions, which alters the regulatory sensitivity \(f'_{NM}\) and, consequently, the feedback gain \(\mathcal{H}\).

Specifically, a value of \(\delta > 1\) indicates strong feedback from \(Y \dashrightarrow X\) while \(X \dashrightarrow Y\) is relatively weak. In contrast, when \(\delta < 1\), it signifies weak feedback but stronger forward regulation. Therefore, by varying \(\delta\), we can investigate how these unequal regulatory strengths affect the cyclic contributions in different TNF motifs.

The time-scale asymmetry is defined as $A_\tau = \beta_X/\beta_Y$. When \( A_\tau = 1 \), it indicates that the degradation timescales of the two TFs are equal. If \( A_\tau < 1 \), it means that TF \( X \) relaxes more slowly than TF \( Y \). Conversely, if \( A_\tau > 1 \), TF \( X \) relaxes more rapidly than TF \( Y \).  Since $\mathcal{T}_{MN}$ is influenced by the degradation rates, altering $A_{\tau}$ changes the time-averaging factor associated with the cyclic noise. Together, the parameters $\delta$ and $A_{\tau}$ enable us to distinguish the effects of regulatory asymmetry and timescale asymmetry on cyclic noise in specific motifs. In Appendix Figures \ref{fig_a2} and \ref{fig_a3}, we illustrate the behavior of intrinsic and extrinsic noise contributions as functions of \( \delta \) and \( A_\tau \).

Fig.~\ref{fig2}a-d shows an intriguing behavior of the cyclic noise for both TFs $X$ and $Y$ in all TNF variants as a function of $\delta$. In the $(++)$ TNF motif, increasing $\delta$ redistributes the cyclic contribution between the two nodes: $\eta_{\text{cyc},X}^2$ decreases, whereas $\eta_{\text{cyc},Y}^2$ increases as the feedback activation $Y\rightarrow X$ becomes stronger (Fig.~\ref{fig2}a). The two cyclic components of $X$ and $Y$ become equal at $\delta=1$ due to regulatory symmetry. Thus, moving from the weak- to the strong-feedback regime shifts the cyclic contribution from $X$ to $Y$. This is because the strong positive feedback regulates $X$ precisely, and thereby decreases its intrinsic noise ($\eta_{\text{int},X}^2$). In this regime, forward regulation becomes weaker, thereby increasing the intrinsic noise of $Y$ ($\eta_{\text{int},Y}^2$). However, in the $(--)$ TNF, the cyclic noise shows a non-monotonic dependence on $\delta$, with node-specific maxima (Fig.~\ref{fig2}b). As $\delta$ increases, $\eta_{\text{cyc},X}^2$ first rises due to a gradual increase in feedback repression strength. However, in a very strong feedback repression regime, the forward regulation becomes very weak, and the loop-closure effect is reduced, which yields a reduction in $\eta_{\text{cyc},X}^2$. The same mechanism applies to $Y$, but with the maximum shifted to the opposite side of $\delta=1$, because the forward repression $X\dashv Y$ is stronger in the weak-feedback regime. For the $(-+)$ TNF, as $\delta$ increases, the feedback activation $Y\rightarrow X$ becomes stronger, whereas the forward repression $X\dashv Y$ becomes weaker. Since the cyclic contribution depends on the combined effect of both regulatory arms, weakening the forward repression reduces the effective loop-closure contribution. As a result, the negative cyclic noise becomes smaller in magnitude as the system moves toward the strong-feedback regime (Fig.~\ref{fig2}c). The $(+-)$ motif represents the mirror image of the $(-+)$ motif, and hence, it shows a mirrored behavior with respect to $\delta$ (Fig.~\ref{fig2}d).

Fig.~\ref{fig2}e-h shows the dependence of cyclic noise on $A_\tau$ at fixed $\delta=1$. In the $(++)$ TNF motif, both $\eta_{\text{cyc},X}^2$ and $\eta_{\text{cyc},Y}^2$ show a non-monotonic dependence on $A_\tau$ (Fig.~\ref{fig2}e). However, the two cyclic components become comparable near $A_\tau=1$, where the relaxation times are balanced. Thus, changing $A_\tau$ produces node-specific maxima while redistributing the cyclic contribution from $X$ to $Y$. In the $(--)$ TNF motif, a similar non-monotonic dependence is observed, but the maxima are more sharply separated between the two nodes (Fig.~\ref{fig2}f). For the $(-+)$ TNF motif, $\eta_{\text{cyc},X}^2$ moves steadily toward zero as $A_\tau$ increases, whereas $\eta_{\text{cyc},Y}^2$ shows a weak non-monotonic variation before approaching zero at larger $A_\tau$ (Fig.~\ref{fig2}g). Therefore, the cyclic contribution becomes progressively weaker in the fast-$X$ regime. The $(+-)$ motif shows a mirrored behavior with respect to $A_\tau$ (Fig.~\ref{fig2}h).

These results show that cyclic noise is strongly shaped by both feedback strength and timescale asymmetry. While $\delta$ tunes the imbalance between the two regulatory arms of the loop, $A_\tau$ changes how the cyclic contribution is weighted by the relative relaxation dynamics of $X$ and $Y$. Thus, cyclic noise provides a motif-specific readout of how feedback architecture and dynamical asymmetry jointly shape stochastic fluctuations in TNF motifs.


\begin{figure*}[!t]
    \centering
    \includegraphics[width=1\linewidth]{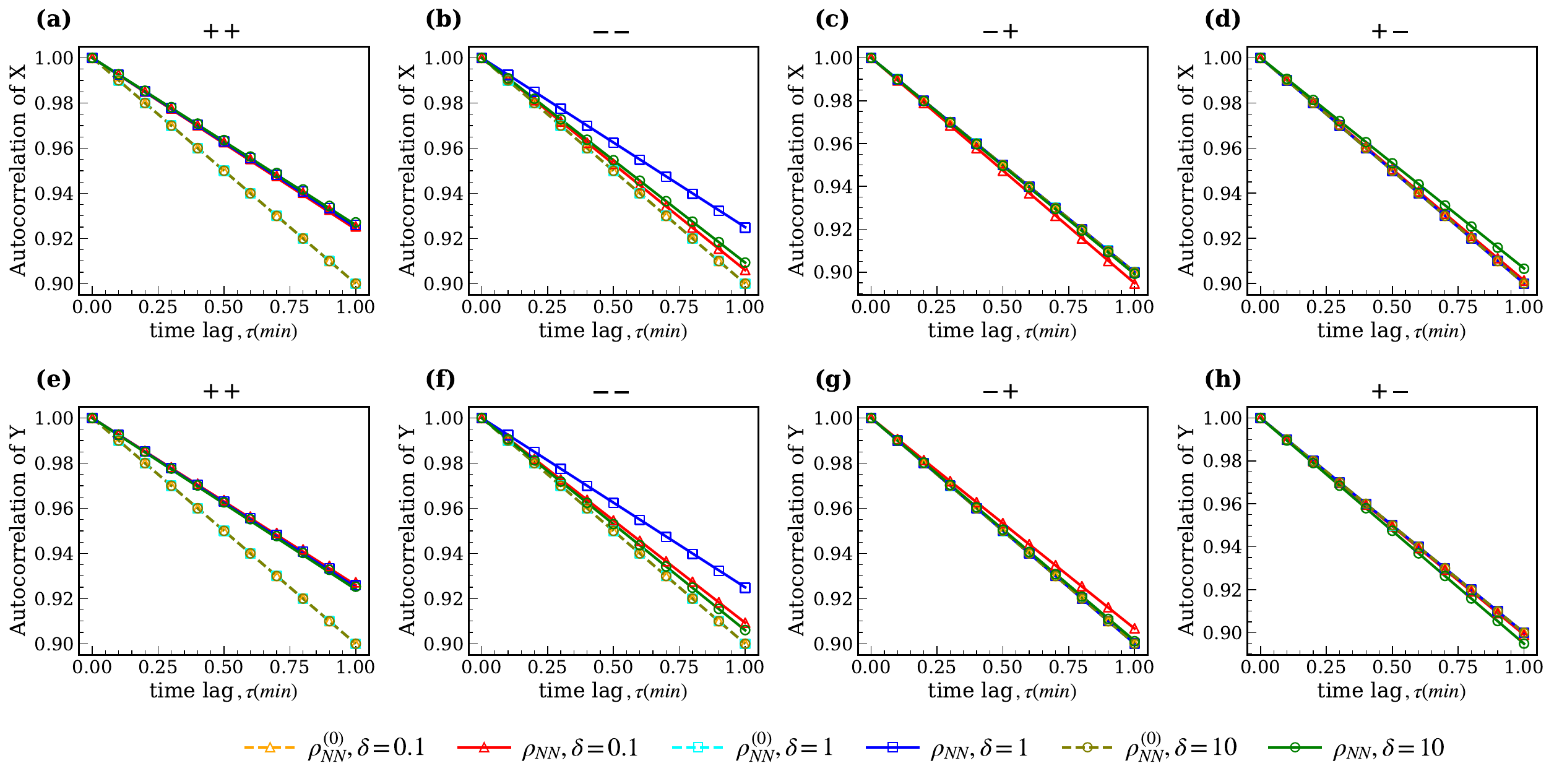}
    \caption{Dependence of steady-state autocorrelation on feedback strength $\delta$ in TNF motifs. 
    Autocorrelation profiles are shown as functions of time lag $\tau$ for three values of $\delta$, at fixed $A_\tau=1$. 
    Panels (a)-(d) show autocorrelations of TF $X$, whereas panels (e)-(h) show autocorrelations of TF $Y$. 
    Panels (a,e), (b,f), (c,g), and (d,h) correspond to $(++)$, $(--)$, $(-+)$, and $(+-)$ TNF motifs, respectively.
    The quantitie $\rho_{NN}^{(0)}$ denotes autocorrelation for no-feedback references, whereas $\rho_{NN}$ denotes autocorrelation in the feedback-coupled TNF motifs, where $N \in \{X,Y\}$. 
    Solid and dashed lines represent theoretical predictions, whereas symbols denote stochastic simulation results averaged over $10^5$ trajectories \cite{Gillespie1976, Gillespie1977}. 
    The parameters are $\alpha_X=10$ (molecules/V)min$^{-1}$, $\alpha_Y=10$ (molecules/V)min$^{-1}$, $\beta_X=0.1\sqrt{A_\tau}$ min$^{-1}$, $\beta_Y=0.1/\sqrt{A_\tau}$ min$^{-1}$, $A_\tau=1$, $K_{XY}=50\sqrt{\delta}$ molecules/V, and $K_{YX}=50/\sqrt{\delta}$ molecules/V.
    }
    \label{fig3}
\end{figure*}

\subsection{Temporal signature of feedback-mediated noise circulation}

The decomposition of steady-state noise reveals how fluctuations are partitioned into intrinsic, extrinsic, and cyclic components. However, this noise decomposition does not describe how long these fluctuations persist. As the cyclic noise reflects a feedback-mediated circulation of fluctuations, this process should leave a temporal signature in the decay of steady-state fluctuations. We therefore ask whether feedback loop closure modifies the temporal persistence of fluctuations in a topology-dependent manner. Specifically, we compare the autocorrelation of each TNF motif with its corresponding no-feedback relaxation profile to determine whether loop closure prolongs, weakens, or leaves the persistence of steady-state fluctuations nearly unchanged. We quantify this temporal persistence using the steady-state autocorrelation function.

To quantify this effect, we compute the steady-state autocorrelation function under the LNA. The time-lagged covariance matrix of the fluctuations $\bm{\delta n}(t)=(\delta x(t),\delta y(t))^\top$ is defined as,
\begin{equation}
    \mathbf{C}(\tau) = \langle \bm{\delta n}(t+\tau) \bm{\delta n}(t)^\top \rangle, \quad \tau \ge 0.
    \label{eq8}
\end{equation}

\noindent Under the LNA of Eq.~(\ref{eq1}), we derive this time-lagged covariance matrix at short time lag (see Appendix~\ref{a3}), given as,
\begin{equation}
    \mathbf{C}(\tau) = e^{\mathbf{J}\tau} \mathbf{\Sigma} \approx (\mathbf{I}+\mathbf{J}\tau)\mathbf{\Sigma},
    \label{eq9}
\end{equation}

\noindent where $\mathbf{J}$ and $\mathbf{\Sigma}$ are the Jacobian and covariance matrix, respectively [see Eq.~(\ref{eq2})]. Here, $\mathbf{I}$ stands for the identity matrix. While deriving Eq.~(\ref{eq9}), we assume that $\lvert \lambda_i \rvert \tau \ll 1$, where $\lambda_i$ ($i \in \{ 1,2 \}$) is the $i$th eigenvalue of the Jacobian $\mathbf{J}$. We note that this short-lag expansion allows us to separate the leading birth-death relaxation of each TF from the additional contribution produced by feedback. It can therefore reveal how loop closure initially changes the decay of steady-state fluctuations. In the full matrix exponential, these contributions are mixed with higher-order coupling effects and cannot be separated explicitly in this form.

\begin{figure*}[!t]
    \centering
    \includegraphics[width=1\linewidth]{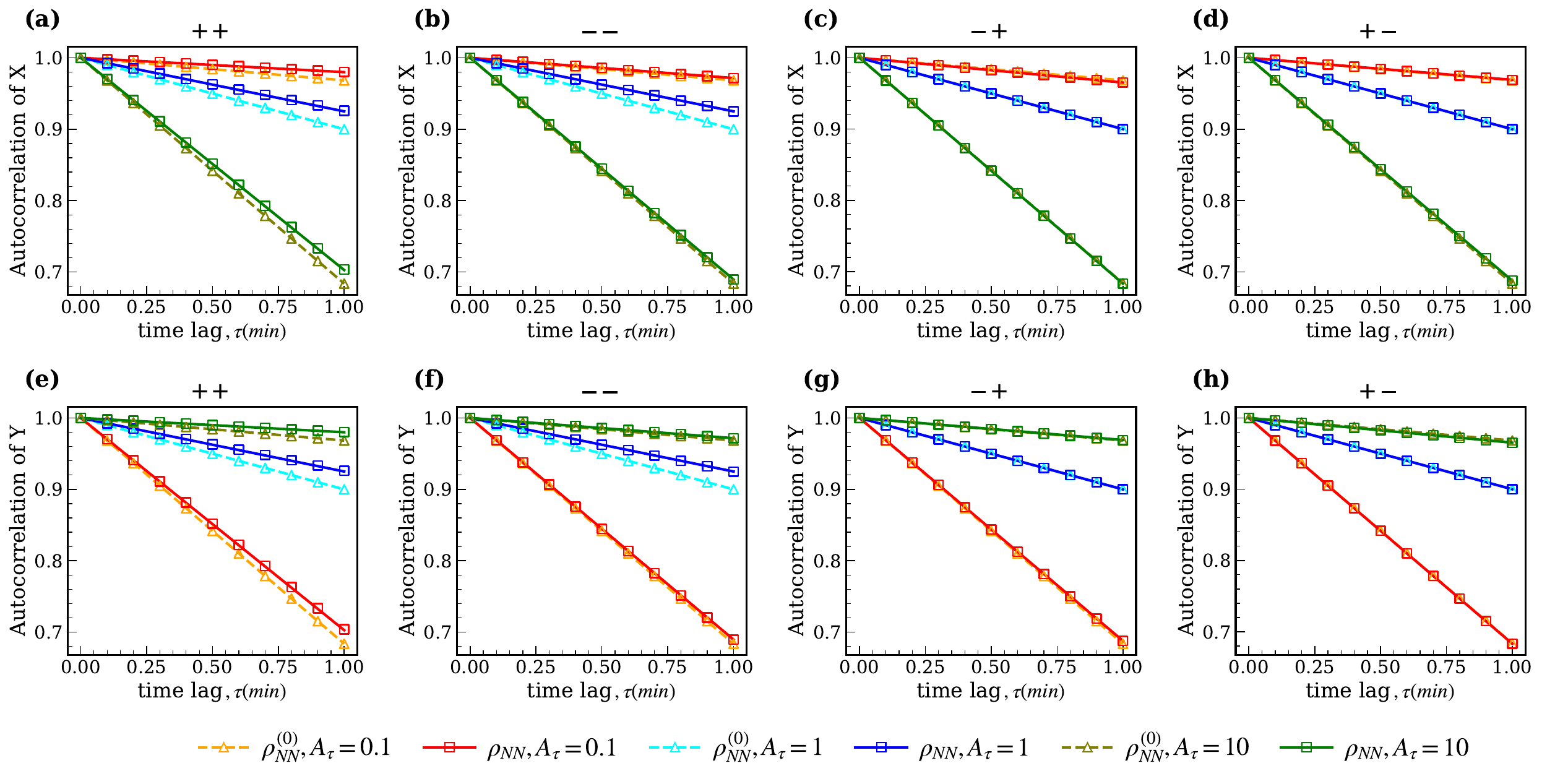}
    \caption{Dependence of steady-state autocorrelation on timescale asymmetry $A_\tau$ in TNF motifs. 
    Autocorrelation profiles are shown as functions of time lag $\tau$ for three values of $A_\tau$, at fixed $\delta=1$. 
    Panels (a)-(d) show autocorrelations of TF $X$, whereas panels (e)-(f) show autocorrelations of TF $Y$. 
    Panels (a,e), (b,f), (c,g), and (d,h) correspond to $(++)$, $(--)$, $(-+)$, and $(+-)$ TNF motifs, respectively.
    The quantitie $\rho_{NN}^{(0)}$ denotes autocorrelation for no-feedback references, whereas $\rho_{NN}$ denotes autocorrelation in the feedback-coupled TNF motifs, where $N \in \{X,Y\}$. 
    Solid and dashed lines represent theoretical predictions, whereas symbols denote stochastic simulation results averaged over $10^5$ trajectories \cite{Gillespie1976, Gillespie1977}. 
    The parameters are $\alpha_X=10$ (molecules/V)min$^{-1}$, $\alpha_Y=10$ (molecules/V)min$^{-1}$, $\beta_X=0.1\sqrt{A_\tau}$ min$^{-1}$, $\beta_Y=0.1/\sqrt{A_\tau}$ min$^{-1}$, $K_{XY}=50\sqrt{\delta}$ molecules/V, $K_{YX}=50/\sqrt{\delta}$ molecules/V, and $\delta=1$.
    }
    \label{fig4}
\end{figure*}

The normalized autocorrelation of TF $N$ is then given by the diagonal terms of $\mathbf{C}(\tau)$, as,
\begin{eqnarray}
    \rho_{NN}(\tau) = \frac{C_{NN}(\tau)}{\sigma_N^2} = \rho_{NN}^{(0)}(\tau) + \hat{\rho}_{NN}(\tau),
    \label{eq10}
\end{eqnarray}

\noindent where the first term $\rho_{NN}^{(0)}(\tau) = 1-\beta_N\,\tau$ is the leading short-lag relaxation associated with the birth-death dynamics of $N$. This term is the first-order approximation to the relaxation $e^{-\beta_N \tau}$ and can appear in the absence of feedback. However, the feedback-mediated coupling contribution is captured by the second term defined as,
\begin{equation}
    \hat{\rho}_{NN}(\tau) = \tau f_{NM}^\prime \left(\frac{\langle n_M\rangle}{\langle n_N\rangle}\right) \frac{\eta_{MN}^2}{\eta_N^2},
\end{equation}

\noindent where, $\eta_{MN}^2$ is the normalized covariance between $M$ and $N$ ($N \neq M$). In essence, $\hat{\rho}_{NN}(\tau)$ captures both the effect of extrinsic noise propagation and cyclic noise in the TNF motifs. We compare $\rho_{NN}(\tau)$ and $\rho_{NN}^{(0)}(\tau)$ for all TNF motifs for different values of $\delta$ and $A_\tau$ (see Figs.~\ref{fig3},\ref{fig4}).


Fig.~\ref{fig3} shows the dependence of autocorrelation on the feedback strength $\delta$ at fixed $A_\tau=1$. In all TNF motifs, the autocorrelation starts from unity at zero lag and decays with increasing time lag, indicating the gradual loss of self-correlation of the initial fluctuation. In the reinforcing motifs, $(++)$ and $(--)$, $\rho_{NN}(\tau)$ clearly deviates from the no-feedback reference $\rho_{NN}^{(0)}(\tau)$ over a finite time window (Fig.~\ref{fig3}a,b,e,f). This indicates that feedback loop closure prolongs the temporal persistence of fluctuations. However, unlike the cyclic-noise amplitude, the autocorrelation profiles are nearly insensitive to changes in $\delta$ for $(++)$ motif, but are sensitive for $(--)$ motif. For the opposing motifs, $(-+)$ and $(+-)$, the profiles of $\rho_{NN}(\tau)$ remain closer to the no-feedback profiles, $\rho_{NN}^{(0)}(\tau)$ (Fig.~\ref{fig3}c,d,g,h). This indicates that opposing feedback produces no significant persistence effect. Thus, changing $\delta$ modifies the temporal persistence of fluctuations in a motif-specific way.

Fig.~\ref{fig4} shows the dependence of autocorrelation on the timescale asymmetry $A_\tau$ at fixed $\delta=1$. In contrast to the $\delta$-dependent case, changing $A_\tau$ produces a significant change in the autocorrelation profiles because it directly changes the relaxation rates of the two TFs. For $X$, the autocorrelation $\rho_{XX}(\tau)$ persists longest when $A_\tau=0.1$, becomes intermediate at $A_\tau=1$, and decays fastest when $A_\tau=10$ (Fig.~\ref{fig4}a-d). This ordering follows the relaxation of $X$: smaller $A_\tau$ corresponds to slower relaxation of $X$, whereas larger $A_\tau$ corresponds to faster relaxation. For $Y$, the opposite ordering is observed. The autocorrelation of $Y$, $\rho_{YY}(\tau)$, decays fastest at $A_\tau=0.1$, becomes intermediate at $A_\tau=1$, and persists longest at $A_\tau=10$ (Fig.~\ref{fig4}e-h). Thus, $A_\tau$ creates an opposite redistribution of temporal persistence between the two nodes.

The comparison with $\rho_{NN}^{(0)}(\tau)$ shows how feedback modifies this relaxation-controlled baseline. In the reinforcing motifs, $(++)$ and $(--)$, the feedback-mediated autocorrelation generally decays more slowly than the corresponding no-feedback reference (Fig.~\ref{fig4}a,b,e,f). Thus, reinforcing loop closure enhances the temporal persistence of fluctuations most clearly. For the opposing motifs, $(-+)$ and $(+-)$, the autocorrelation $\rho_{NN}(\tau)$ remains close to the no-feedback reference $\rho_{NN}^{(0)}(\tau)$ (Fig.~\ref{fig4}c,d,g,h). Thus, $A_\tau$ primarily determines the baseline duration of autocorrelation, whereas feedback topology determines whether loop closure prolongs or leaves this persistence unchanged.

These autocorrelation analyses show that feedback topology controls the temporal persistence of fluctuations in TNF motifs. The reinforcing motifs, $(++)$ and $(--)$, show a clearer and longer-lived deviation of the autocorrelation from the no-feedback reference, indicating that loop closure prolongs the temporal persistence of fluctuations. In contrast, the opposing motifs, $(-+)$ and $(+-)$, remain mostly closer to the no-feedback relaxation profiles. This distinction suggests that positive cyclic noise is associated not only with noise amplification but also with longer temporal persistence, whereas negative cyclic noise is associated with noise attenuation and mostly unaltered autocorrelation.


\section{Conclusion}

This study develops a theoretical framework to understand how feedback topology shapes noise propagation in TNF motifs. By applying the LNA to a general stochastic model of mutually regulating TFs, the total noise of each node is decomposed into intrinsic, extrinsic, and cyclic components. The central contribution of this framework is the cyclic noise term, which captures the feedback-dependent contribution generated by loop closure. Unlike intrinsic and extrinsic noise, it is positive for reinforcing feedback motifs, $(++)$ and $(--)$, and negative for opposing feedback motifs, $(-+)$ and $(+-)$. Thus, the framework provides an analytical way to distinguish feedback-mediated noise amplification from feedback-mediated noise attenuation in gene-regulatory circuits. The analysis further shows that cyclic noise is not determined solely by the feedback topology. Feedback strength, quantified by $\delta$, and timescale asymmetry, quantified by $A_\tau$, tune the magnitude and node-wise distribution of the cyclic contribution. In addition, the autocorrelation analysis provides a temporal counterpart to feedback-mediated noise propagation. Reinforcing motifs prolong the persistence of fluctuations, whereas opposing motifs produce no significant change relative to the no-feedback relaxation. Therefore, feedback topology controls not only the amplitude of steady-state fluctuations but also the duration over which these fluctuations remain temporally correlated.

We note that a previous study by Kobayashi \textit{et al.} \cite{Kobayashi2016} accounted for the feedback-loop closure effect through the overall source-decomposition structure, rather than a separate and explicit additive noise component. In contrast, our work moves beyond this formulation by isolating the loop-closure effect as a distinct node-wise cyclic noise term alongside the intrinsic and extrinsic components. This explicit separation allows us to compare how feedback topology amplifies or attenuates fluctuations across the four TNF motifs. Furthermore, the autocorrelation analysis adds a temporal layer by showing how feedback-mediated circulation affects the persistence of fluctuations, thereby extending the analysis beyond steady-state noise decomposition.

This framework may be relevant for interpreting noise regulation in feedback-containing gene-regulatory systems, including cell-fate decision circuits, developmental switches, and synthetic gene circuits \cite{Alon2007, Losick2008, Eldar2010, Gardner2000}. In such systems, the sign and magnitude of cyclic noise suggest a possible way to distinguish whether feedback is expected to amplify variability, buffer fluctuations, or redistribute noise between interacting regulators. Since the theory is expressed in terms of measurable quantities such as mean copy numbers, variances, covariance, and autocorrelation, it may also provide a basis for future comparisons of feedback architectures using single-cell time-series data.

The present framework is formulated within the LNA, which limits the analysis to small fluctuations and does not address strongly nonlinear regimes such as rare switching events, transcriptional bursting, explicit delays, or large-amplitude oscillations. However, this restriction also provides the main strength of our approach: it allows the total noise to be decomposed analytically into intrinsic, extrinsic, and cyclic components, and it connects each contribution directly to measurable quantities such as mean copy numbers, variances, covariances, and autocorrelations. Thus, rather than providing a complete description of all possible feedback dynamics, the present theory establishes a minimal and interpretable framework for identifying feedback-mediated noise circulation in two-node regulatory motifs. Future extensions can build on this baseline by incorporating bursting, delays, non-Gaussian fluctuations, and larger regulatory networks.

The present framework identifies cyclic noise as an analytically tractable signature of feedback-mediated circulated fluctuation. The main implication is that feedback does not merely change the magnitude of noise; it also determines whether fluctuations are amplified, attenuated, or temporally retained through loop closure. This minimal framework, therefore, provides a basis for future extensions to larger regulatory networks and for experimental tests using single-cell dynamics.

\section*{Acknowledgments}

N.R. thanks the Bose Institute, Kolkata, for the Junior Research Fellowship.


\appendix

\renewcommand{\thefigure}{A\arabic{figure}} 
\setcounter{figure}{0}


\section{Moments associated with $X$ and $Y$}
\label{a1}

The steady-state Jacobian matrix $\mathbf{J}$, covariance matrix $\mathbf{\Sigma}$, and diffusion matrix $\bm{\mathcal{D}}$ for the TNF motifs are given by,
\begin{eqnarray*}
\mathbf{J} &=& \begin{pmatrix}
- \beta_{X} & f^\prime_{XY} \\
f^\prime_{YX} & -\beta_{Y}
\end{pmatrix},
\quad
\mathbf{\Sigma} = \begin{pmatrix}
\sigma_X^2 & \sigma_{XY}^2 \\
 \sigma_{YX}^2 & \sigma_Y^2
\end{pmatrix},
\\
\bm{\mathcal{D}} &=& 
\begin{pmatrix}
\frac{f_X(\langle x \rangle, \langle y \rangle) + \beta_X \langle x \rangle}{2} & 0 \\
0 & 
\frac{f_Y(\langle x \rangle, \langle y \rangle) + \beta_Y \langle y \rangle}{2}
\end{pmatrix}
\\
& = &
\begin{pmatrix}
\beta_X \langle x \rangle & 0 \\
0 & \beta_Y \langle y \rangle
\end{pmatrix}.
\end{eqnarray*}

\noindent In defining $\mathbf{J}$, we use $f^\prime_{XY} = [\partial f_X(x, y)/\partial y]_{\bm{n}=\langle \bm{n}\rangle}$ and  $f^\prime_{YX} = [\partial f_Y(x, y)/\partial x]_{\bm{n}=\langle \bm{n}\rangle}$, which are the steady-state regulatory sensitivities for $Y \dashrightarrow X$ and $X \dashrightarrow Y$ regulations, respectively. In defining $\bm{\mathcal{D}}$, we have used the steady-state condition $f_N(\langle \bm{n}\rangle)=\beta_N \langle n_N\rangle$ due to Eq.~(\ref{eq1}). On solving the Lyapunov equation [Eq.~(\ref{eq2})], we get the analytical expressions for elements of the covariance matrix $\mathbf{\Sigma}$,
\begin{eqnarray}
\sigma^{2}_{X} &=& \langle x \rangle \left[1 + \frac{f^\prime_{XY} f'_{YX}\, \beta_{Y}}{(\beta_{X}+ \beta_{Y})(- f^\prime_{XY} f^\prime_{YX} + \beta_{X} \beta_{Y})} \right] \nonumber \\
&& +  \frac{(f^\prime_{XY})^{2} \beta_{Y}\langle y \rangle }{(\beta_{X}+ \beta_{Y})(- f^\prime_{XY} f^\prime_{YX} + \beta_{X} \beta_{Y})}, 
\label{eqa11} \\
\sigma^{2}_{Y} &=& \langle y \rangle \left[1+ \frac{f^\prime_{XY} f^\prime_{YX} \,\beta_{X}}{(\beta_{X}+ \beta_{Y})(- f^\prime_{XY} f^\prime_{YX}  + \beta_{X} \beta_{Y})} \right]  \nonumber \\
&&+ \frac{(f^\prime_{YX})^{2} \beta_{X}\langle x \rangle }{(\beta_{X}+ \beta_{Y})(- f^\prime_{XY} f^\prime_{YX} + \beta_{X} \beta_{Y})},
\label{eqa12} \\
\sigma^{2}_{XY} &=& \sigma_{YX}^2 = \frac{(f^\prime_{YX} \langle x \rangle + f^\prime_{XY} \langle y \rangle)\beta_{X}\beta_{Y}}{(\beta_{X}+ \beta_{Y})(- f^\prime_{XY} f^\prime_{YX} + \beta_{X} \beta_{Y})}.
\label{eqa13}
\end{eqnarray}


\begin{figure}[!t]
    \centering
    \includegraphics[width=1.0\linewidth]{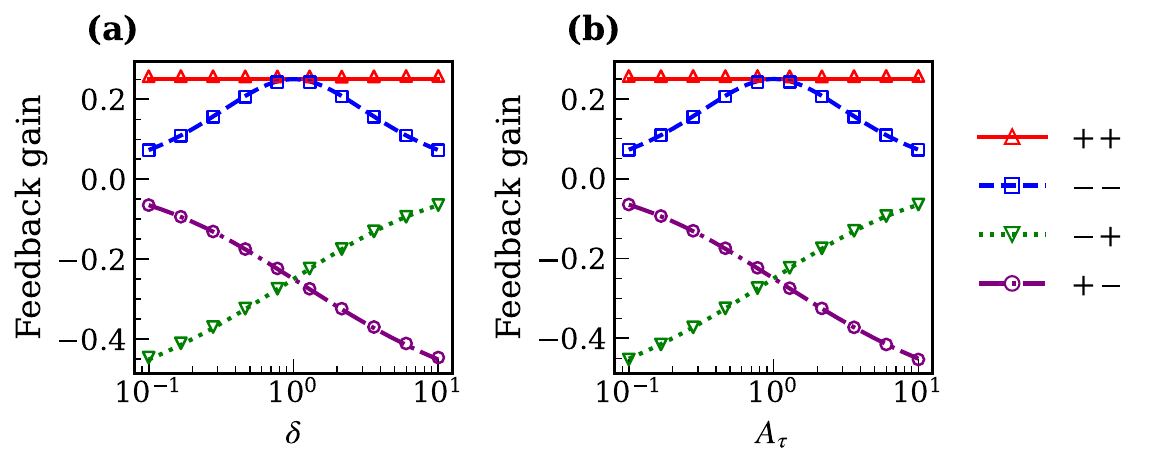}
    \caption{Dependence of feedback gain $\mathcal{H}$ on feedback strength $\delta$ and timescale asymmetry $A_\tau$ in TNF motifs. 
    (a) Variation of feedback gain with $\delta$, at fixed $A_\tau=1$. 
    (b) Variation of feedback gain with $A_\tau$, at fixed $\delta=1$.
    Lines represent theoretical predictions, whereas symbols denote stochastic simulation results averaged over $10^5$ trajectories \cite{Gillespie1976,Gillespie1977}. 
    The parameters are $\alpha_X=10$ (molecules/V)min$^{-1}$, $\alpha_Y=10$ (molecules/V)min$^{-1}$, $\beta_X=0.1\sqrt{A_\tau}$ min$^{-1}$, $\beta_Y=0.1/\sqrt{A_\tau}$ min$^{-1}$, $K_{XY}=50\sqrt{\delta}$ molecules/V, and $K_{YX}=50/\sqrt{\delta}$ molecules/V.
    }
\label{fig_a1}
\end{figure}


\section{Noise decomposition}
\label{a2}

The noise associated with each node of the TNF motif is quantified by squared CV, $\eta_N^2=\sigma_N^2/\langle n_N\rangle^2$. Using Eqs.~(\ref{eqa11}-\ref{eqa12}), we write the noise decomposition for TFs $X$ and $Y$ as,
\begin{eqnarray}
\eta_X^2 &=& \frac{\sigma_X^2}{{\langle x \rangle}^2}
= \eta_{\text{int},X}^2 + \eta_{\text{ext},X}^2 + \eta_{\text{cyc},X}^2, \\
\eta_{Y}^2 &=& \frac{\sigma_Y^2}{{\langle y \rangle}^2}
=\eta_{\text{int},Y}^2 + \eta_{\text{ext},Y}^2 + \eta_{\text{cyc},Y}^2.
\end{eqnarray}

\noindent The explicit expressions of intrinsic noise ($\eta_{\text{int},N}^2$), extrinsic noise ($\eta_{\text{ext},N}^2$) and cyclic noise ($\eta_{\text{cyc},N}^2$) are,
\begin{eqnarray}
\eta_{\text{int},X}^2 &=& \frac{1}{\langle x \rangle} \\
\eta_{\text{int},Y}^2 &=& \frac{1}{\langle y \rangle} \\
\eta_{\text{ext},X}^2 &=& \frac{(f^\prime _{XY})^2 \langle y \rangle}{\beta_{X} (\beta_{X} +\beta_{Y})\langle x \rangle} \, \eta_{\text{int},X}^2 \, \frac{1}{1- \mathcal{H}}\\
\eta_{\text{ext},Y}^2 &=& \frac{(f^\prime_{YX})^2 \langle x \rangle}{\beta_{Y} (\beta_{X} +\beta_{Y})\langle y \rangle} \, \eta_{\text{int},Y}^2 \, \frac{1}{1-\mathcal{H}}\\
\eta_{\text{cyc},X}^2 &=& \mathcal{T}_{YX} \, \, \frac{\mathcal{H}}{1- \mathcal{H}} \, \, \eta_{\text{int},X}^2 \\
\eta_{\text{cyc},Y}^2 &=& \mathcal{T}_{XY} \, \, \frac{\mathcal{H}}{1- \mathcal{H}} \, \, \eta_{\text{int},Y}^2,
\end{eqnarray}

\noindent where, $\mathcal{H}= (f^\prime_{XY} \, f^\prime_{YX})/(\beta_X \, \beta_Y)$ is defined as the feedback gain, and $\mathcal{T}_{XY}=\beta_{X}/(\beta_X +\beta_Y)$ and $\mathcal{T}_{YX}=\beta_Y/(\beta_X+\beta_Y)$ are the time-averaging factors.


\begin{figure*}[!t]
    \centering
    \includegraphics[width=1\linewidth]{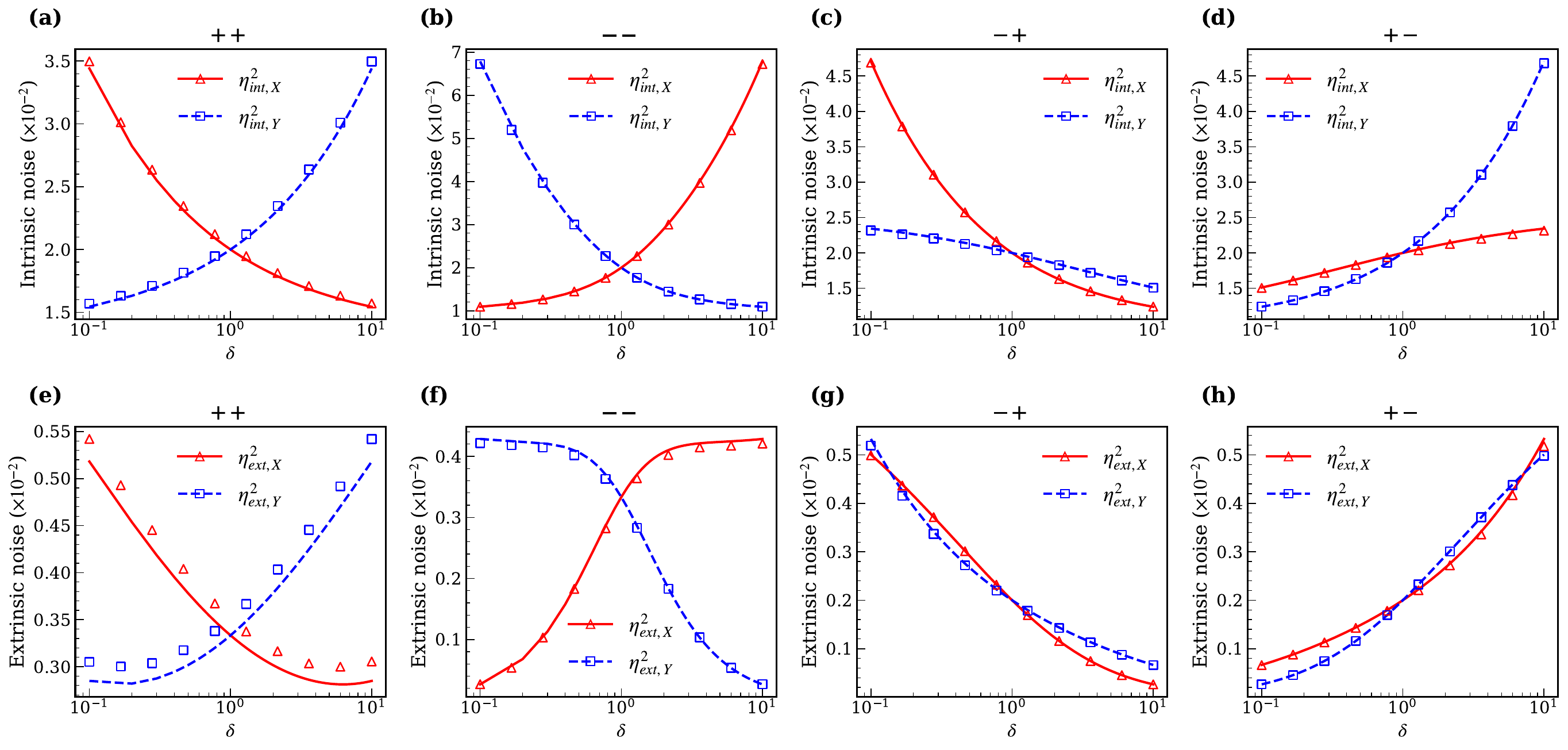}
    \caption{Dependence of intrinsic and extrinsic noise on feedback strength $\delta$ in TNF motifs. 
    (a--d) Variation of intrinsic noise with $\delta$, at fixed $A_\tau=1$. 
    (e--h) Variation of extrinsic noise with $\delta$, at fixed $A_\tau=1$. 
    Panels (a,e), (b,f), (c,g), and (d,h) correspond to $(++)$, $(--)$, $(-+)$, and $(+-)$ TNF motifs, respectively.
    Lines represent theoretical predictions, whereas symbols denote stochastic simulation results averaged over $10^5$ trajectories \cite{Gillespie1976,Gillespie1977}. 
    The parameters are $\alpha_X=10$ (molecules/V)min$^{-1}$, $\alpha_Y=10$ (molecules/V)min$^{-1}$, $\beta_X=0.1\sqrt{A_\tau}$ min$^{-1}$, $\beta_Y=0.1/\sqrt{A_\tau}$ min$^{-1}$, $A_\tau=1$, $K_{XY}=50\sqrt{\delta}$ molecules/V, and $K_{YX}=50/\sqrt{\delta}$ molecules/V.
    }
        \label{fig_a2}
\end{figure*}


\begin{figure*}[!t]
    \centering
    \includegraphics[width=1\linewidth]{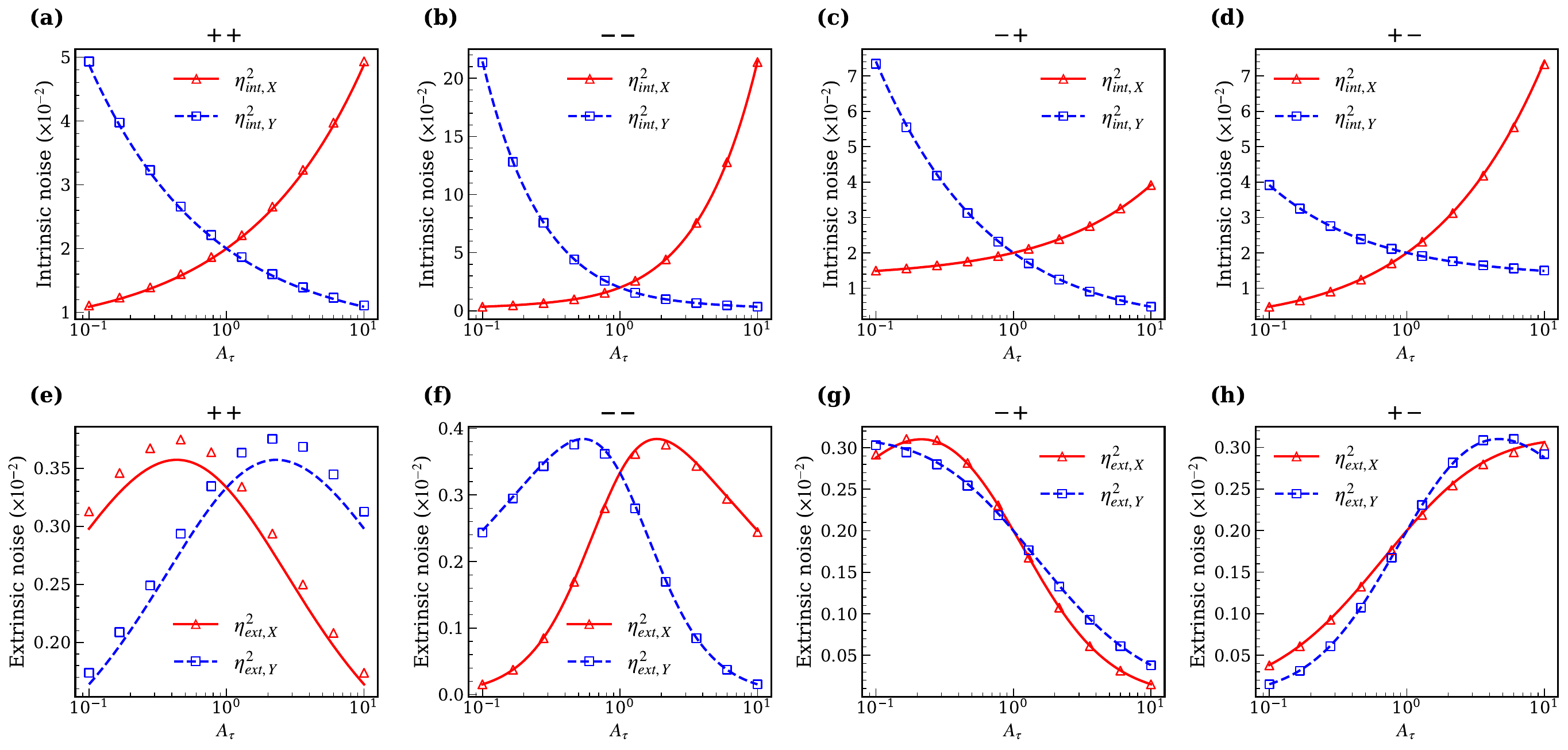}
    \caption{Dependence of intrinsic and extrinsic noise on timescale asymmetry $A_\tau$ in TNF motifs. 
    (a--d) Variation of intrinsic noise with $A_\tau$, at fixed $\delta=1$. 
    (e--h) Variation of extrinsic noise with $A_\tau$, at fixed $\delta=1$. 
    Panels (a,e), (b,f), (c,g), and (d,h) correspond to $(++)$, $(--)$, $(-+)$, and $(+-)$ TNF motifs, respectively.
    Lines represent theoretical predictions, whereas symbols denote stochastic simulation results averaged over $10^5$ trajectories \cite{Gillespie1976, Gillespie1977}. 
    The parameters are $\alpha_X=10$ (molecules/V)min$^{-1}$, $\alpha_Y=10$ (molecules/V)min$^{-1}$, $\beta_X=0.1\sqrt{A_\tau}$ min$^{-1}$, $\beta_Y=0.1/\sqrt{A_\tau}$ min$^{-1}$, $K_{XY}=50\sqrt{\delta}$ molecules/V, $K_{YX}=50/\sqrt{\delta}$ molecules/V, and $\delta=1$.
    }
    \label{fig_a3}
\end{figure*}


\section{Derivation of the time-lagged covariance and autocorrelation functions}
\label{a3}

Under the LNA, the fluctuations in copy number can be defined as $\bm{\delta n}(t)=\bm{n}(t) - \langle \bm{n}\rangle \equiv (\delta n_X, \delta n_Y)^\top$. The dynamics of $\bm{\delta n}(t)$ are derived from Eq.~(\ref{eq1}), yielding a linearized Langevin equation of the form \cite{Kampen2007},
\begin{equation}
    \frac{d \bm{\delta n}(t)}{dt} = \mathbf{J} \bm{\delta n}(t) + \sqrt{2}\mathbf{\Delta} \bm{\xi}(t),
    \label{eqa31}
\end{equation}

\noindent Equivalently, this equation can be written as a stochastic differential equation,
\begin{equation}
    d\bm{\delta n}(t) = \mathbf{J}\bm{\delta n}(t)dt + \sqrt{2}\mathbf{\Delta}\,d\mathbf{W}(t),
    \label{eqa32}
\end{equation}

\noindent where $\mathbf{W}(t)$ is a standard Wiener process satisfying $d\mathbf{W}(t)/dt=\bm{\xi}(t)$, $\langle d\mathbf{W}(t)\rangle=\mathbf{0}$, and $\langle d\mathbf{W}(t)d\mathbf{W}(t)^\top\rangle=\mathbf{I}\,dt$ \cite{Gillespie2000, Erban2020}. Here, $\mathbf{0}$ and $\mathbf{I}$ stand for the zero vector and identity matrix. To calculate the temporal correlation of fluctuations, we define the time-lagged covariance matrix at steady-state as $\mathbf{C}(\tau) = \left\langle \bm{\delta n}(t+\tau)\bm{\delta n}(t)^\top\right\rangle$, for $\tau\geq 0$. The formal solution of Eq.~(\ref{eqa32}) within the time-window $t$ to $t+\tau$ is
\begin{equation}
    d\bm{\delta n}(t+\tau) = e^{\mathbf{J}\tau} d\bm{\delta n}(t) + \sqrt{2} \int_t^{t+\tau} e^{\mathbf{J}(t+\tau-s)} \mathbf{\Delta}\,d\mathbf{W}(s).
    \label{eqa33}
\end{equation}

\noindent Multiplying both sides by $\bm{\delta n}(t)^\top$ and then averaging over the steady-state ensemble gives,
\begin{widetext}
\begin{eqnarray}
    \left\langle
    \bm{\delta n}(t+\tau)\bm{\delta n}(t)^\top \right\rangle & = &
    e^{\mathbf{J}\tau} \left\langle \bm{\delta n}(t)\bm{\delta n}(t)^\top \right\rangle +
    \sqrt{2} \left\langle \int_t^{t+\tau} e^{\mathbf{J}(t+\tau-s)} \mathbf{\Delta}\,d\mathbf{W}(s) \bm{\delta n}(t)^\top
    \right\rangle.
    \label{eqa34}
\end{eqnarray}
\end{widetext}

The second term vanishes because the Wiener increments in the interval $[t,t+\tau]$ are independent of the fluctuation $\bm{\delta n}(t)$. Therefore,
\begin{equation}
    \mathbf{C}(\tau) = e^{\mathbf{J}\tau} \mathbf{\Sigma},
    \label{eqa35}
\end{equation}

\noindent where the covariance matrix $\mathbf{\Sigma} =: \langle \bm{\delta n}(t)\bm{\delta n}(t)^\top\rangle$. For a short time lag, the matrix exponential can be expressed as $e^{\mathbf{J}\tau}=\mathbf{I}+\mathbf{J}\tau+\mathcal{O}(\tau^2)$ and Eq.~(\ref{eqa35}) becomes,
\begin{equation}
    \mathbf{C}(\tau) \simeq \left( \mathbf{I}+\mathbf{J}\tau\right) \mathbf{\Sigma}.
    \label{eqa36}
\end{equation}

\noindent We note that while expanding the exponential $e^{\mathbf{J}\tau}$, we assume that $\lvert \lambda_i \rvert \tau \ll 1$, where $\lambda_i$ ($i \in \{ 1,2 \}$) is the $i$th eigenvalue of the Jacobian $\mathbf{J}$. 

Using the elements of $\mathbf{J}$ and $\mathbf{\Sigma}$, the diagonal elements of $\mathbf{C}(\tau)$ are,
\begin{eqnarray}
    C_{XX}(\tau) &=& \sigma_X^2 + \tau \left(-\beta_X\sigma_X^2 + f'_{XY}\sigma_{XY}^2 \right),
    \label{eqa37} \\
    C_{YY}(\tau) &=& \sigma_Y^2 + \tau \left(- \beta_Y\sigma_Y^2 + f'_{YX}\sigma_{XY}^2 \right).
    \label{eqa38}
\end{eqnarray}

The normalized autocorrelation functions are defined as $\rho_{XX}(\tau)=C_{XX}(\tau)/\sigma_X^2$ and $\rho_{YY}(\tau)=C_{YY}(\tau)/\sigma_Y^2$. Therefore,
\begin{eqnarray}
    \rho_{XX}(\tau) &=& 1-\beta_X\tau + \tau f'_{XY} \left(\frac{\sigma_{XY}^2} {\sigma_X^2} \right) 
    \nonumber \\
    & \equiv & 1-\beta_X\tau + \tau f'_{XY} \frac{\langle y\rangle} {\langle x\rangle} \frac{\eta_{XY}^2} {\eta_X^2},
    \label{eqa39}
    \\
    \rho_{YY}(\tau) &=& 1-\beta_Y\tau + \tau f'_{YX} \left(\frac{\sigma_{XY}^2} {\sigma_Y^2} \right) 
    \nonumber \\
    & \equiv & 1-\beta_Y\tau + \tau f'_{YX} \frac{\langle x\rangle} {\langle y\rangle} \frac{\eta_{XY}^2} {\eta_Y^2},
\end{eqnarray}

\noindent where the normalized covariance $\eta_{XY}^2=\sigma_{XY}^2/\langle x\rangle \langle y\rangle$.



%

\end{document}